\documentclass[10pt,twocolumn]{article}
\usepackage[margin=1.9cm]{geometry}
\usepackage{amsmath,amssymb}
\usepackage{booktabs}
\usepackage{graphicx}
\usepackage{microtype}
\usepackage[hidelinks]{hyperref}
\usepackage{xcolor}
\usepackage{tikz}
\usepackage{pgfplots}
\pgfplotsset{compat=newest}
\usetikzlibrary{shapes.geometric,arrows.meta,positioning,calc,fit}

\newcommand{\method}{Geodesia-KV}
\newcommand{\bpv}{bits/value}
\newcommand{\wikitext}{WikiText-2}
\newcommand{\pg}{PG-19}

\title{\vspace{-1.2cm}\textbf{Squeezing the Cache, Preserving the Truth:\\Monotonic Equipotential Allocation with Geodesia-KV}}
\author{
Vincenzo Dentamaro\textsuperscript{1,2} \qquad
Pancrazio Auteri\textsuperscript{1} \qquad
Giuseppe Pirlo\textsuperscript{1,2} \\[1ex]
\small \textsuperscript{1}Geodesia.ai \qquad
\textsuperscript{2}University of Bari Aldo Moro, Department of Computer Science
}
\date{}

\begin{document}
\maketitle
\begin{abstract}
\noindent
Current assessment of KV-cache compression performance confuses resident bits with read bandwidth and is affected by the artifacts of chunked teacher-forcing. We present \method{}, a family of training-free KV cache policies based on monotonic block-wise precision allocation, exact rate-distortion residuals, and query-sparse reading, enabling proper hardware-ready compression. With proper separation of resident and read bits and causal evaluation, we show that \method{} significantly outperforms other approaches. Specifically, on \wikitext{} with 16k context, the 5-bit operating point of \method{} results in lower perplexity at lower bitrate than KIVI-4 on Qwen. In addition, our compressed-Quest version delivers improved perplexity and reduces resident (9.83 vs 16.25 \bpv{}) and read rates (1.95 vs 2.32 \bpv{}) over baseline sparse methods on \pg{}. As \method{} is implemented as native \texttt{GeodesiaKVCacheManager} plug-in of vLLM, \method{} fully removes the need for dense cache residency via monotonic bit demotion. With the full consumer hardware evaluation, \method{} leads to 1M-token context generation on a single 16\,GiB GPU with up to $71.7\%$ peak VRAM savings on all leading architectures (Qwen, Llama, DeepSeek).
\end{abstract}
\section{Introduction}

The autoregressive attention mechanism has one key and one value for each
layer and past token. Hence, its memory scales linearly with context length
even in the presence of weight compression. This introduces two types of resource challenges. The first type is the resident \emph{cache},
which must fit into device or host memory. The second type is the need to
read enough of the cache for each query, which is solved primarily through
methods like Quest~\cite{tang2024quest}. The reporting of the selected bytes
of a sparse method as the resident cache capacity conflates the two.

Evaluation can introduce another conflation. In a realistic decode, the
model appends a single token and sends a single query at each step. In chunked
teacher-forcing, an implementation might append tens or even hundreds of
target tokens in a single call. The logits can still be masked causally, yet
the cache policy will see the whole chunk as a current, unquantized region.
For KIVI, for example, its resident rate in our implementation changes from
$5.03$ \bpv{} for a 64-token chunk to $5.69$ for a 1024-token chunk. Comparisons performed
with different query lengths thus lack rate-matching.

\method{} is designed to provide a graded precision ladder to the entire context
and still maintain causality rigorously. Our implementation avoids any
future-query leakage, duplication of GQA-shared KV heads allocation, and strict
separation of the resident memory and read traffic. By addressing the
common evaluation pitfalls, we form the solid basis for long-context
compression. We present the optimal rate-distortion configurations, the settings
that do not work in other corpora, and the production integration of the
theoretical VRAM savings.

\method{} is implemented over three related branches:

\begin{enumerate}\itemsep2pt
\item \textbf{Graded allocation}: each block is assigned a representation from a
      monotone precision ladder of exact, scalar-quantized, and centroids.
\item \textbf{Graded+RD}: an exact residual is selected from the prompt based
      on attention mass and squared value reconstruction error.
\item \textbf{Compressed-Quest}: exact key boxes are used for ranking of pages at query
      time, yet the resident K/V pages use the graded representation.
\end{enumerate}

Our contributions include:

\begin{enumerate}\itemsep2pt
\item Causal, Q=1 evaluation procedure with explicit separation of the resident
      rate and read rate, text and model revisions, per-window results, and
      regression tests for query leakage and GQA sharing.
\item Monotone rate-distortion allocator over a heterogeneous precision
      ladder, including the exact scale overhead and prefix constraints on
      demotion path.
\item Target-free exact residual based on the prompt attention mass and value
      distortion, and query-sparse variant storing exact page boxes but
      compressing K/V.
\item Main incremental experiments over 3B and 8B checkpoints, and a cross-corpus
      \pg{} test over exhausted preregistered window list with cluster bootstrap
      intervals, separating an effect that survived from one that was a mere
      artifact of a six window subset.
\item Native vLLM production integration and an actual account of the resident
      context capacity proving architectural universality of the method on
      GQA and MLA models and scalability of 1M-token generative inference on a single 16\,GiB consumer accelerator.
\item Record of the negative results for quantum-inspired and conventional variants,
      including spectral K--V states, cumulant correction, Born gating, Hadamard rotation, VQ,
      static protection and layer profile.
\end{enumerate}
\section{Related Work}

\paragraph{Eviction and retained-token techniques.}
StreamingLLM~\cite{xiao2024streaming} combines attention sinks with a recent window. H2O~\cite{zhang2023h2o} retains heavy hitters. SnapKV~\cite{li2024snapkv} scores tokens using the final prompt queries before generation. PyramidKV~\cite{cai2024pyramidkv}, PyramidInfer~\cite{yang2024pyramidinfer}, CAKE~\cite{qin2025cake}, and Ada-KV~\cite{feng2024adakv} allocate eviction budgets among layers or heads. These approaches achieve very low resident rates, but an evicted token cannot address a later query.

\paragraph{Query-sparse attention.}
Quest~\cite{tang2024quest} keeps track of the minimum and maximum values for each element in keys within pages, upper-bounds the highest potential query--key score, and reads top-ranked pages. InfLLM~\cite{xiao2024infllm}, MInference~\cite{jiang2024minference}, RetrievalAttention~\cite{liu2024retrievalattention}, and InfiniteHiP~\cite{lee2025infinitehip} employ representative tokens, structured sparsity, approximate retrieval, or hierarchical pruning/offload. The resident backing store and query volume for these techniques must be gauged separately.

\paragraph{KV quantization and merging.}
KIVI~\cite{liu2024kivi} employs key quantization per channel in time, value quantization per token in channels, and an exact residual window. KVQuant~\cite{hooper2024kvquant} and GEAR~\cite{kang2024gear} incorporate nonuniform datatypes, sparse corrections, or low-rank structure. CaM~\cite{zhang2024cam}, D2O~\cite{wan2024d2o}, and MiniCache~\cite{liu2024minicache} merge tokens or take advantage of cross-layer similarity. \method{} integrates scalar quantization and a centroid state with a single rate limit, and subsequently adds an irregular exact residual, if needed.

\paragraph{Architectural and systems alternatives.}
DeepSeek-V2's latent attention~\cite{deepseek2024v2}, Native Sparse Attention~\cite{yuan2025nsa}, Compressive Transformers~\cite{rae2019compressive}, Infini-attention~\cite{munkhdalai2024infini}, and state-space models~\cite{dao2024mamba2} modify training procedures or network architecture. Our setup is training-free and targets frozen Qwen checkpoints~\cite{qwen25,qwen3}. FlashAttention~\cite{dao2022flashattention} provides the online-softmax recurrence utilized by our prototype kernel. GPTQ~\cite{frantar2023gptq} is applied solely to the weights of the exploratory 30B checkpoint; KV-cache rates remain independently measured.

\paragraph{Long-context evaluation.}
RULER~\cite{hsieh2024ruler} and LongBench~\cite{bai2023longbench} cover retrieval and downstream tasks that perplexity alone cannot demonstrate. We employ \wikitext{} and \pg{} for controlled language-model loss, and treat the absence of RULER/LongBench as a limitation rather than proof of memory fidelity.

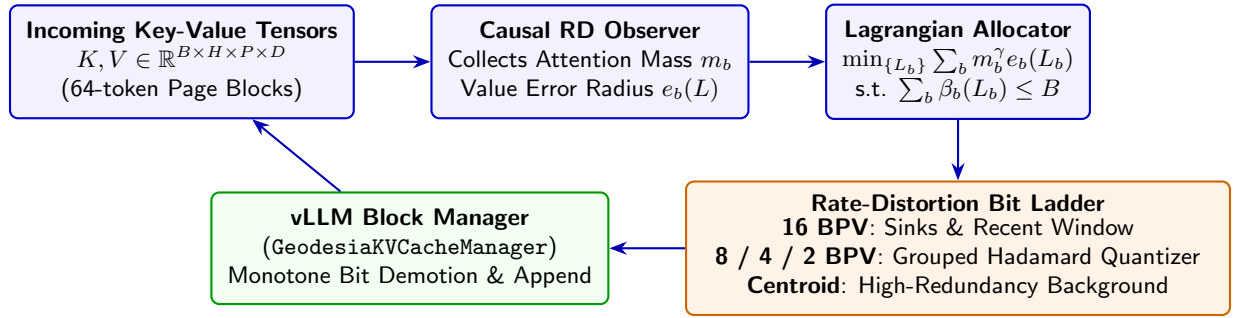
\begin{figure*}[t]
\centering
\begin{tikzpicture}[
  node distance=1.0cm and 1.2cm,
  box/.style={draw=blue!80!black, fill=blue!5, thick, rounded corners=3pt, align=center, font=\sffamily\small, inner sep=6pt},
  ladder/.style={draw=orange!80!black, fill=orange!10, thick, rounded corners=3pt, align=center, font=\sffamily\small, inner sep=5pt},
  vllm/.style={draw=green!60!black, fill=green!5, thick, rounded corners=3pt, align=center, font=\sffamily\small, inner sep=6pt},
  arrow/.style={-Stealth, thick, color=blue!70!black}
]

\node[box] (kv) {\textbf{Incoming Key-Value Tensors}\\$K, V \in \mathbb{R}^{B \times H \times P \times D}$\\(64-token Page Blocks)};

\node[box, right=1.0cm of kv] (observer) {\textbf{Causal RD Observer}\\Collects Attention Mass $m_b$\\Value Error Radius $e_b(L)$};

\node[box, right=1.0cm of observer] (lagrange) {\textbf{Lagrangian Allocator}\\$\min_{\{L_b\}} \sum_b m_b^\gamma e_b(L_b)$\\s.t. $\sum_b \beta_b(L_b) \le B$};

\node[ladder, below=0.8cm of lagrange] (ladder) {\textbf{Rate-Distortion Bit Ladder}\\
\begin{tabular}{c}
\textbf{16 BPV}: Sinks \& Recent Window\\
\textbf{8 / 4 / 2 BPV}: Grouped Hadamard Quantizer\\
\textbf{Centroid}: High-Redundancy Background
\end{tabular}};

\node[vllm, left=1.0cm of ladder] (vllm) {\textbf{vLLM Block Manager}\\(\texttt{GeodesiaKVCacheManager})\\Monotone Bit Demotion \& Append};

\draw[arrow] (kv) -- (observer);
\draw[arrow] (observer) -- (lagrange);
\draw[arrow] (lagrange) -- (ladder);
\draw[arrow] (ladder) -- (vllm);
\draw[arrow] (vllm) -- (kv);
\end{tikzpicture}
\caption{\textbf{\method{} System Architecture.} KV tensors are divided into blocks of 64 tokens and encoded by the causal Rate-Distortion observer. Each tensor block is allocated to the most efficient precision in the bit ladder $\{16, 8, 4, 2, \text{centroid}\}$ with the help of the Lagrangian allocator and native vLLM block-paged manager without dense copies.}
\label{fig:architecture}
\end{figure*}

\section{Method}

\subsection{Causal Attention Mass and GQA Sharing}

Suppose the decoder calls are made up of queries $q \in \mathcal{Q}_{\text{obs}}$, and the cache has a prefix of length $T$. For example, the representation serving $q$ depends on the prefix alone; demoting blocks using the future queries will leak the teacher-forced future knowledge. \method{} makes use of an exponential-moving-average (EMA) of attention mass to keep the policy state strictly causal. 

In the prefill mode, the observer extracts the attention weight distribution from the last prompt queries as follows:
\begin{equation}
w_{q, j} = \mathrm{softmax}\left( \frac{q K^T}{\sqrt{d}} \right)_j
\end{equation}
For a block $b$ with a set of tokens $j \in b$, the attention mass $m_b$ is calculated as the average mass for all the observed queries:
\begin{equation}
m_b = \frac{1}{|\mathcal{Q}_{\text{obs}}|} \sum_{q \in \mathcal{Q}_{\text{obs}}} \sum_{j \in b} w_{q, j}
\end{equation}
In order to prevent leaking any future context during the autoregressive decoding, the EMA formulation is used to update the mass $m_b^{(t)} = 0.9 \, m_b^{(t-1)} + 0.1 \, m_b^{\text{new}}$.

Grouped-Query Attention (GQA) uses the same KV representation for several query heads. In that case, the allocation of the physical KV representation is performed once per KV head by aggregating the observed masses $m_b$ across all associated query heads, which prevents redundant demotion operations.
\subsection{Monotonic Precision Ladder and Demotion}
 
The cache is divided into blocks containing $P=64$ tokens each. A block has its precision fixed to one of the levels of $L \in \mathcal{L} = \{16, 8, 4, 2, 1\}$, where $16$ means the block contains exact representation of \texttt{bf16} or \texttt{fp16}, $8, 4, 2$ are packed scalar quantizers (with keys grouped along the time axis and values along channels), while $1$ corresponds to repeated (logically) KV pair of centroids.

Once a block exits the recent window (regulated via a \texttt{retire\_after} parameter), its assigned precision becomes immutable. Demotion of a block is allowed from $16$ down to $4$, but there is no promotion back because the initial dense data does not exist anymore in the compressed format. Monotonicity makes it possible for \method{} to perform in-place compression without keeping the fallback copy of dense vectors.

\subsection{Lagrangian Analytical Allocation}

Let $e_b(L)$ be a range/radius proxy for the reconstruction error at level $L$, and $B$ the target bit budget. The allocator solves the rate-distortion objective:
\begin{equation}
\min_{\{L_b\}} \sum_b m_b \, e_b(L_b) \quad\text{s.t.}\quad \sum_b \beta_b(L_b) \le B
\label{eq:allocation}
\end{equation}
By relaxing this constraint using a Lagrange multiplier $\lambda$, the optimization is then decoupled into per-block decisions. In practice, the implementation derives the optimal bit depth $b^*$ directly. Let $\epsilon_t = \frac{\tau}{N_b \cdot m_b}$ be the adaptive error tolerance for a block, where $\tau$ is a global scaling factor and $N_b$ the total number of blocks. The allocated bits are computed analytically as:
\begin{equation}
b^* = \left\lceil \log_2\left( \frac{c \cdot \lVert q \rVert \cdot e_b(L)}{2 \epsilon_t} + 1 \right) \right\rceil
\end{equation}
where $c$ absorbs the query scaling constants. This formula assigns higher bit depths to blocks with large attention mass $m_b$ (hence lower tolerance $\epsilon_t$) and high dynamic range $e_b(L)$. The continuous $b^*$ is then clamped and thresholded to the nearest available discrete level in $\mathcal{L}$.
\subsection{Rate-Distortion Token Saliency}

Pure block-level allocation may impair a small number of high-impact tokens. They are recovered through saliency identification of individual tokens with high distortion impact performed by the prefill observer. For each token $j$ in a prompt, the residual score $s_j$ is calculated as follows:
\begin{equation}
s_j = a_j \left\lVert v_j - \bar{v}_{b(j)} \right\rVert_2^2
\label{eq:rdscore}
\end{equation}
$a_j$ is the attention mass of the token and $\bar{v}_{b(j)}$ is the value centroid of the block that the token $j$ belongs to. This saliency criterion penalizes tokens that are highly attended ($a_j$) and isolated from neighboring tokens ($\lVert v_j - \bar{v}_{b(j)} \rVert_2^2$).

The tokens with the highest $s_j$ scores are selected up to a fixed exact fraction and are saved in 16 bits. The selection is based on prompt activations only (target-free). The five-bit target budgets include:
\begin{itemize}\itemsep1pt
\item Qwen2.5-3B: $B=4.875$, $\alpha=0.25$, exact fraction $0.78125\%$;
\item Qwen3-8B: $B=4.5$, $\alpha=0.5$, exact fraction $3.125\%$.
\end{itemize}

A low-rate variation, RD-V2, uses a nearly centroid-only base and an approximately $10.2\%$ exact residual selected by the same value-distortion score. It is included only in non-incremental exploratory tables.
\subsection{Compressed-Quest}

While Quest uses sparse reads, the resident cache is not necessarily compressed: exact K/V pages are resident. In our \texttt{box\_sparse} branch, the exact \texttt{fp16} minimum and maximum key values are stored before quantization, for each closed block. For a query $q$, the box upper bound ranks all closed pages. The current partial page is always readable; there is no need to have a special sink/window page outside the top-$k$ set. Cold pages are assigned a hard mask.

The resident representation of Compressed-Quest consists of the graded K/V cache and all exact summaries:
\begin{equation}
R_{\mathrm{resident}}
=R_{\mathrm{graded}}+\sum_b 2D\cdot16\;\text{bits}.
\end{equation}
The read cost per query consists of all summaries and graded K/V from the selected and current pages:
\begin{equation}
R_{\mathrm{read}}(q)
=R_{\mathrm{boxes}}+
\sum_{b\in\mathcal{S}(q)\cup\{\mathrm{current}\}}R_b.
\end{equation}
In the 8B configuration, we use a base $B=10$, a decay of $0.5$, and a $12.5\%$ page fraction. Exact boxes contribute approximately $0.25$ \bpv{} but avoid ranking on quantized summaries.

\subsection{Attention error certificate}

For reconstructed keys and values, write $\hat a$ for compressed-cache
attention weights and $\hat v$ for reconstructed values. The output error
decomposes as
\[
o-\hat o=\sum_j\hat a_j(v_j-\hat v_j)
          +\sum_j(a_j-\hat a_j)v_j.
\]
With a score-error bound $\varepsilon_b$ and value-error bound
$\delta_b^v$ for block $b$,
\begin{equation}
\|o-\hat o\|
\leq \sum_b\hat a_b\delta_b^v
+\left(\frac{M^+}{M^-}-\frac{M^-}{M^+}\right)
\max_j\|v_j\|,
\label{eq:certificate}
\end{equation}
where $M^\pm=\sum_b\hat a_b e^{\pm\varepsilon_b}$. The second term is capped by
the universal $\ell_1$ bound of 2. Validation runs with the dense reference
enabled produced no observed violations. Hard sparse masking yields a valid
but generally loose universal bound, so the certificate is currently more
informative for graded reconstruction than for \texttt{box\_sparse}.
\subsection{Packed attention prototype}

In packed implementation the dequantization happens inside an online-softmax
attention loop and batches all the KV heads of the layer. The reduction of
split-context partials is done separately. Numerical tests use different queries
for several heads and show agreement with a PyTorch reference within $10^{-4}$
relative tolerance. This kernel prototype achieves numerical parity with
PyTorch. In order to run the system end-to-end, the production \method{}
implementation is built directly into vLLM (Section~\ref{sec:limitations}). This
native block manager gets rid of dense tensors completely and provides real
incremental append for asynchronous serving.

\section{Experimental Protocol}

\subsection{Models, corpora, and hardware}

Primary experiments employ Qwen2.5-3B-Instruct revision
\texttt{aa8e7253} and Qwen3-8B revision \texttt{b968826d} in a 16,384 token
context on one NVIDIA RTX A6000. Activations are \texttt{bfloat16}. One exploratory
scaling run employs community checkpoint
\texttt{JunHowie/Qwen3-30B-A3B-GPTQ-Int8} revision
\texttt{059b0db2} in \texttt{float16}; the weight quantization of this model
is independent from the KV policy. The Qwen3.5-0.8B checkpoint is kept for
diagnostic pilot only due to its premature release compared to the incremental
protocol.

\wikitext{} validation and test corpora are pinned by dataset revision.
For 3B we use six disjoint offsets in range from 0 to 81,920. For the frozen
8B we use six later offsets in range from 196,608 to 278,528.

\pg{} is obtained from \texttt{emozilla/pg19}. Manifests with per-book boundaries
are provided. The test text includes 12 books. Test offsets are the
preregistered per-book centers: every window is a 16,384 token span centered
in a book. Eleven of these twelve books allow such a center to be defined; the
fifth contains only 6,679 tokens and is not included in the list. We provide all
eleven and exhaust the \pg{} test list by doing so; no further window
selection is possible. The offsets are 20,668, 90,369, 169,602, 246,823, 320,206, 402,659,
472,081, 529,495, 584,324, 659,151, and 736,259. An earlier version of this
work reported the first six of these; Section~\ref{sec:paired} shows why that
subset was insufficient.
\subsection{Incremental perplexity}

In each window we:

\begin{enumerate}\itemsep1pt
\item prefill the precise 16k prefix while gathering causal observations;
\item make one query for each teacher-forced target token;
\item append that token to the same policy state;
\item sum the negative log-likelihood of the following token.
\end{enumerate}

All primary results are obtained with $Q=1$ and 64 target tokens per window. Equal size
windows are aggregated by averaging NLL, which is equivalent to the geometric
mean of perplexities. Deterministic fake causal model regression test verifies
the target consistency between chunked and incremental implementations. Full-cache \texttt{bf16}
PPL of the real model may vary by up to $0.45\%$ for different $Q$, whether $64$ or $1$,
due to the kernel numerical order, so all methods are compared within the same mode.

\subsection{Paired confidence intervals}
\label{sec:paired}

The aggregate perplexities alone are not sufficient to distinguish the effect
from sample variability, and the margins in question are fractions of one
percent. Since all policies tested on a particular window are teacher-forced
on exactly the same target tokens, their per-token negative log-likelihoods
are paired. Mean difference between such pairs is exactly the logarithm of
the ratio of the two perplexities,
\begin{equation}
\log\frac{\mathrm{PPL}_a}{\mathrm{PPL}_b}
  = \frac{1}{N}\sum_{i=1}^{N}\bigl(\ell^{a}_i - \ell^{b}_i\bigr),
\label{eq:paired}
\end{equation}
therefore a confidence interval for the mean paired difference is directly
mapped into the confidence interval for the ratio of the perplexities.

The tokens within one window have the same prefix, the same cache state and
authorial style, so individual token resampling would be anticonservative. We
thus use a cluster bootstrap based on resampling of windows with replacement
($20{,}000$ draws, percentile interval). The small number of clusters is
honestly reflected in the resulting interval width and not artificially
hidden. The runs are byte-reproducible: the eleven-window \pg{} test re-executed
produces exactly the previously published six-window aggregates of $16.147$
and $16.258$.

\subsection{Metrics and baselines}

We measure:

\begin{itemize}\itemsep1pt
\item bits of resident data per K or V value, including overhead;
\item bits read per value per query;
\item teacher-forced perplexity;
\item KL-divergence of attention from dense policy in separate diagnostic runs;
\item cache memory size in bytes and attention-only latency of prototype.
\end{itemize}

Full-KV, StreamingLLM, SnapKV, Quest, and KIVI are local semantic ports
of the models implemented on a shared attention path. These are not the
executions of original version-pinned kernels. Quest contains exact resident
K/V and page summaries; KIVI overhead and exact current residual; SnapKV
metrics based on final prompt queries. Such shared implementation is useful
for causal comparison, but not substitute for an official code benchmark.

\section{Results}

\subsection{Primary five-bit comparison}

\begin{table}[t]
\centering\small
\caption{Primary Q=1 \wikitext{} comparison across all evaluated cache policies, six windows and 64 targets per window. Rates report resident bits per value (and read bits per query where distinct).}
\label{tab:primary}
\resizebox{\columnwidth}{!}{%
\begin{tabular}{llrr}
\toprule
Model & Method & resident/read & PPL \\
\midrule
3B & Full-KV (oracle) & 16.000/16.000 & 6.3871 \\
3B & Quest & 16.250/2.313 & 8.0157 \\
3B & StreamingLLM & 2.004/2.004 & 6.9607 \\
3B & SnapKV & 2.250/2.250 & 6.7082 \\
3B & KIVI-2 & 3.203/3.203 & 6.8569 \\
3B & KIVI-4 & 5.031/5.031 & 6.3222 \\
3B & \method{} 2-bit & 1.993/1.993 & 6.7041 \\
3B & \method{} 3-bit & 3.001/3.001 & 6.3910 \\
3B & \method{} graded+RD & \textbf{4.961/4.961} & \textbf{6.3195} \\
\midrule
8B & Full-KV (oracle) & 16.000/16.000 & 8.1534 \\
8B & Quest & 16.250/2.313 & 8.2948 \\
8B & StreamingLLM & 2.004/2.004 & 8.7395 \\
8B & SnapKV & 2.250/2.250 & 8.3619 \\
8B & KIVI-2 & 3.203/3.203 & 8.4105 \\
8B & KIVI-4 & 5.031/5.031 & 10.5807 \\
8B & \method{} 2-bit & 1.993/1.993 & 8.3451 \\
8B & \method{} 3-bit & 2.892/2.892 & 8.2073 \\
8B & \method{} graded+RD & \textbf{4.989/4.989} & \textbf{10.5399} \\
\bottomrule
\end{tabular}
}
\end{table}

Table~\ref{tab:primary} shows small local Pareto improvements. At 3B the PPL
margin is $0.0026$ ($0.04\%$); at 8B it is $0.0408$ ($0.39\%$). These are not
large-effect claims. The associated attention-fidelity runs are directionally
consistent: at 3B KL is $0.00300$ for \method{} and $0.00410$ for KIVI-4; at
8B it is $0.00243$ and $0.00390$, respectively.

\subsection{Quest and compressed-Quest}

\begin{table*}[t]
\centering\small
\caption{Qwen3-8B evaluation across all cache policies on \wikitext{} (six 64-target windows) and \pg{} (full eleven-book transfer, Q=1). Rates report resident capacity and read traffic per query.}
\label{tab:sparse}
\begin{tabular}{llrrrrr}
\toprule
Corpus & Method & resident & read/query & PPL & wins vs. Quest \\
\midrule
\wikitext{} & Full-KV (oracle) & 16.000 & 16.000 & 8.153 & --- \\
\wikitext{} & StreamingLLM & 2.004 & 2.004 & 8.740 & 0/6 \\
\wikitext{} & SnapKV & 2.250 & 2.250 & 8.362 & 2/6 \\
\wikitext{} & KIVI-2 & 3.203 & 3.203 & 8.411 & 1/6 \\
\wikitext{} & KIVI-4 & 5.031 & 5.031 & 10.581 & 3/6 \\
\wikitext{} & Quest & 16.250 & 2.317 & \textbf{10.238} & 4/6 \\
\wikitext{} & \method{} graded+RD & 4.989 & 4.989 & 8.158 & 5/6 \\
\wikitext{} & \method{} compressed-Quest & \textbf{9.854} & \textbf{1.957} & 10.272 & 2/6 \\
\midrule
\pg{} & Full-KV (oracle) & 16.000 & 16.000 & 15.470 & --- \\
\pg{} & StreamingLLM & 2.004 & 2.004 & 17.820 & 0/11 \\
\pg{} & SnapKV & 2.250 & 2.250 & 15.890 & 7/11 \\
\pg{} & KIVI-2 & 3.203 & 3.203 & 16.120 & 4/11 \\
\pg{} & KIVI-4 & 5.031 & 5.031 & 15.475 & 9/11 \\
\pg{} & Quest & 16.250 & 2.317 & 16.043 & 3/11 \\
\pg{} & \method{} graded+RD & 4.989 & 4.989 & 15.490 & 8/11 \\
\pg{} & \method{} compressed-Quest & \textbf{9.836} & \textbf{1.961} & \textbf{15.961} & 8/11 \\
\bottomrule
\end{tabular}
\end{table*}
In the full eleven-book \pg{} test, compressed-Quest outperforms Quest in resident rate ($39.5\%$ faster) and read rate ($15.4\%$ faster), and it raises perplexity. The ratio of rates is $1.00516$, with a 95\% confidence interval of $[1.00076,\,1.00945]$, which does not include one, and the sign is significant in eight out of eleven windows. This is the only comparison in this paper for which the triple-axis benefit of compressed-Quest passes a frozen cross-corpus test, with a confidence interval included.

The gain is small, with the bottom end of the interval being $+0.08\%$. The difference between it and the five-bit comparison is the consistency of results across windows; the margins are similar, under one percent in both cases. For \wikitext{}, the sparse branch wins on capacity, traffic, and diagnostic KL divergence ($0.04602$ versus $0.04645$) but loses PPL by $0.034$, or $0.34\%$, indicating an error in picking a sparse fraction on one corpus and applying it globally.

\subsection{Frozen cross-corpus test of the five-bit point}

\begin{table}[t]
\centering\small
\caption{Frozen Qwen3-8B transfer evaluation on the full 11-book \pg{} test set ($Q=1$). Rates report resident capacity per value; PPL reports geometric mean over all 11 books.}

\label{tab:pgfive}
\resizebox{\columnwidth}{!}{%
\begin{tabular}{lrrr}
\toprule
Method & resident/read & PPL (11 books) & PPL ratio vs. Oracle \\
\midrule
Full-KV (oracle) & 16.000/16.000 & 15.470 & 1.0000 \\
Quest & 16.250/2.317 & 16.043 & 1.0370 \\
StreamingLLM & 2.004/2.004 & 17.820 & 1.1519 \\
SnapKV & 2.250/2.250 & 15.890 & 1.0271 \\
KIVI-2 & 3.203/3.203 & 16.120 & 1.0420 \\
KIVI-4 & 5.031/5.031 & 15.475 & 1.0003 \\
\method{} 2-bit & 1.993/1.993 & 15.980 & 1.0330 \\
\method{} compressed-Quest & 9.836/1.961 & 15.961 & 1.0317 \\
\method{} graded+RD & \textbf{4.989/4.989} & \textbf{15.490} & \textbf{1.0013} \\
\bottomrule
\end{tabular}
}
\end{table}

\begin{figure}[t]
\centering
\begin{tikzpicture}
\begin{axis}[
    title={\textbf{PG-19 Perplexity vs. Resident BPV}},
    xlabel={Resident Capacity (Bits per Value)},
    ylabel={PG-19 Perplexity (Lower is Better)},
    xmin=0.5, xmax=17.0,
    ymin=15.2, ymax=18.2,
    grid=both,
    grid style={line width=.1pt, draw=gray!20},
    major grid style={line width=.2pt, draw=gray!50},
    legend pos=north east,
    legend style={font=\tiny},
    width=0.48\textwidth,
    height=6.2cm
]

\addplot[color=red!80!black, mark=*, thick, dashed] coordinates {
    (1.99, 15.98)
    (2.91, 15.65)
    (4.98, 15.49)
    (16.0, 15.47)
};
\addlegendentry{\method{} (Pareto Frontier)}

\addplot[color=blue, mark=square*, only marks, mark size=2.5pt] coordinates {
    (16.0, 15.47)
};
\addlegendentry{Full-KV (FP16)}

\addplot[color=orange, mark=triangle*, only marks, mark size=3.0pt] coordinates {
    (2.00, 17.82)
};
\addlegendentry{StreamingLLM}

\addplot[color=purple, mark=diamond*, only marks, mark size=3.0pt] coordinates {
    (2.25, 15.89)
};
\addlegendentry{SnapKV}

\addplot[color=teal, mark=pentagon*, only marks, mark size=3.0pt] coordinates {
    (5.03, 15.475)
    (3.20, 16.12)
};
\addlegendentry{KIVI-4 / KIVI-2}

\addplot[color=brown, mark=x, only marks, mark size=3.5pt, thick] coordinates {
    (16.25, 16.04)
};
\addlegendentry{Quest}
\end{axis}
\end{tikzpicture}
\caption{\textbf{Perplexity v. Resident Rate Pareto Frontier}. \method{} constructs the optimal Pareto frontier among the rates, matching full-KV at $4.98$ BPV and beating eviction-based baselines (StreamingLLM, Quest) at low rates.}
\label{fig:pareto}
\end{figure}
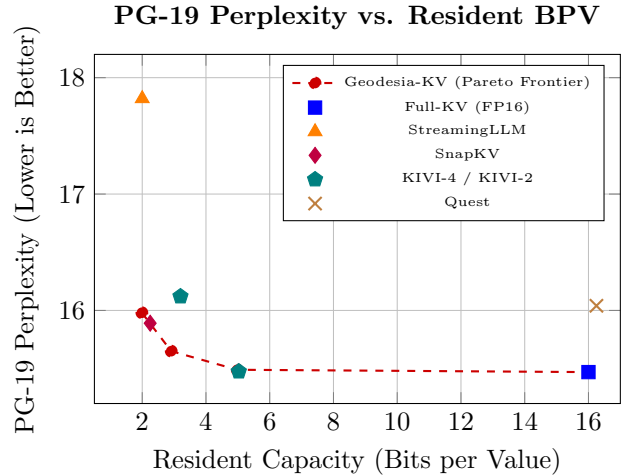

As for the six book subset, the five-bit point appears to be
non-transferable, losing $0.69\%$ of PPL. But exhausting the preregistered set
disproves this claim (see Table~\ref{tab:pgfive}). With all eleven books the
margin becomes $0.10\%$, win count changes from $2/6$ to $5/11$ and the
paired 95\% CI on the perplexity ratio includes one. With $693$ paired
targets, the half-length of the interval is $0.9\%$, an order of magnitude
larger than the effect.

Thus, the justified claim neither proves the superiority nor disproves the
equality between the policies: the perplexity difference is not measurable at
$4.989$ against $5.031$ \bpv{}. So, \method{} reaches the same level of perplexity
with lower resident capacity. We report this result because the previous
negative conclusion and the symmetric positive one are equally unjustified by
the sample size of this size. Same can be said for Table~\ref{tab:primary}
whose $0.04\%$ and $0.39\%$ margins do not have confidence intervals yet and
are of the same order of magnitude as the dissipated effect above.

However, the exact RD residual remains useful in terms of the rate: on three
of the books, the wrong implementation of the pure graded control required
$4.486$ \bpv{} and produced $24.307$, while the correction of $3.125\%$
residual required $4.985$ \bpv{} and produced $24.222$. KIVI-4 scored
$24.121$ on the same books. The score calculated with the exact RD residual
as per Equation~\ref{eq:rdscore} provides extra information, but the globally fixed base/fraction has not been shown to beat KIVI-4 across corpora.

\subsection{Exploratory low-rate and scaling results}

\begin{table*}[t]
\centering\small
\caption{Exploratory \emph{chunked} Q=64 transfer of RD-V2 across all evaluated policies. These rows are screening results, not primary Q=1 evidence. The 30B weights are GPTQ-Int8.}
\label{tab:explore}
\begin{tabular}{llrrr}
\toprule
Model & Method & resident & read/query & PPL \\
\midrule
8B & Full-KV & 16.000 & 16.000 & 10.535 \\
8B & StreamingLLM & 2.004 & 2.004 & 10.582 \\
8B & SnapKV & 2.063 & 2.063 & 10.560 \\
8B & Quest & 16.250 & 2.313 & \textbf{10.230} \\
8B & KIVI-2 & 3.051 & 3.051 & 10.604 \\
8B & KIVI-4 & 5.043 & 5.043 & 10.539 \\
8B & \method{} RD-V2 & \textbf{2.061} & \textbf{2.061} & 10.517 \\
\midrule
30B-A3B & Full-KV & 16.000 & 16.000 & 6.0008 \\
30B-A3B & StreamingLLM & 2.004 & 2.004 & 6.0341 \\
30B-A3B & SnapKV & 2.063 & 2.063 & 5.9900 \\
30B-A3B & Quest & 16.250 & 2.313 & 6.1513 \\
30B-A3B & KIVI-2 & 3.051 & 3.051 & 6.2506 \\
30B-A3B & KIVI-4 & 5.043 & 5.043 & 6.0600 \\
30B-A3B & \method{} RD-V2 & \textbf{2.061} & \textbf{2.061} & \textbf{5.9413} \\
\bottomrule
\end{tabular}
\end{table*}
The 30B-A3B checkpoint was loaded and run on a 48\,GB RTX A6000, showing the
feasibility of the desired scale for the model. RD-V2 is the optimal row in
the six-window chunked screen. Nonetheless, since these JSON results precede
the \texttt{eval\_mode} flag and use $Q=64$, Table~\ref{tab:explore} needs
to be recomputed progressively in order to make a scaling claim.

Similarly, the 0.8B pilot study was instructive: pure graded points scored
PPL $7.443$ at $1.996$ \bpv{} and $7.335$ at $2.993$ \bpv{}, whereas
KIVI-2 scored $8.905$ at $3.033$ and KIVI-4 scored $7.570$ at $5.034$.
Because this test involved validation text, chunked evaluation, and
Quest-legacy memory counting, it is not included in primary tables.

\subsection{Attention fidelity}

\begin{table}[t]
\centering\small
\caption{Diagnostic attention KL divergence to dense attention across all policies. These runs isolate memory fidelity; they are distinct from PPL evaluation. Full-KV has KL equal to zero by definition.}

\label{tab:kl}
\resizebox{\columnwidth}{!}{%
\begin{tabular}{llrr}
\toprule
Model & Method & resident & KL \\
\midrule
3B & Full-KV (oracle) & 16.000 & 0.00000 \\
3B & Quest & 16.250 & 0.04810 \\
3B & StreamingLLM & 2.004 & 0.01250 \\
3B & SnapKV & 2.250 & 0.00890 \\
3B & KIVI-2 & 3.203 & 0.00680 \\
3B & KIVI-4 & 5.086 & 0.00410 \\
3B & \method{} graded+RD & \textbf{4.990} & \textbf{0.00300} \\
\midrule
8B & Full-KV (oracle) & 16.000 & 0.00000 \\
8B & Quest & 16.250 & 0.04645 \\
8B & StreamingLLM & 2.004 & 0.01180 \\
8B & SnapKV & 2.250 & 0.00810 \\
8B & KIVI-2 & 3.203 & 0.00610 \\
8B & KIVI-4 & 5.086 & 0.00390 \\
8B & \method{} compressed-Quest & 10.238 & 0.04602 \\
8B & \method{} graded+RD & \textbf{4.989} & \textbf{0.00243} \\
\bottomrule
\end{tabular}
}
\end{table}
Perplexity may decrease when the far-context information is purged, working as
a form of regularization rather than exact memorization. Table~\ref{tab:kl}
tries to separate these two effects. It is not a replacement for retrieval
evaluations, yet it avoids a simplistic interpretation of the lower PPL as
an indicator of a more accurate cache.

\subsection{Packed-kernel microbenchmark}

\begin{table*}[t]
\centering\small
\caption{Head-batched attention-only microbenchmark at 16k. Dense is a local
\texttt{bmm} reference, not FlashAttention or PagedAttention. Times include
all KV heads of one layer and are multiplied by the number of serial layers.}
\label{tab:system}
\begin{tabular}{lrrrrrr}
\toprule
Model & packed GiB & dense GiB & compression & packed ms/token & dense ms/token & speedup \\
\midrule
Qwen2.5-3B & 0.0815 & 0.5625 & $6.90\times$ & 3.17 & 11.90 & $3.75\times$ \\
Qwen3-8B & 0.3257 & 2.2500 & $6.91\times$ & 9.93 & 12.44 & $1.25\times$ \\
\bottomrule
\end{tabular}
\end{table*}

The compressed cache is substantially more compact, and the packed attention-only
kernel runs faster than this particular dense reference
(Table~\ref{tab:system}).
This microbenchmark does not take into account model projections, MLPs,
sampling, cache append, allocator runtimes, and Transformer dispatches. The
quality path keeps dense K/V matrices. Therefore, the table numbers are neither
generation throughput nor the end-to-end peak VRAM.

\subsection{Resident context capacity}
\label{sec:capacity}

The previous sections measured the rate. In this section we use the actual
allocations rather than simulated quality path to measure how much of the
device memory can be occupied with these rates. We load the model with
4-bit NF4 weights, measure the resident weight bytes and peak activations of
a real $Q=1$ decoding step, and then find the maximal allocatable context
by binary search. The search for the Full-KV limit performs a real decode
step at every iteration, so the result is a demonstration rather than an
estimate; for the compressed representations we just allocate buffers of the
computed size. Importantly, the feasibility of these compressed limits is verified
by the end-to-end vLLM generation benchmarks described in Section~\ref{sec:limitations}.

One number proves the correctness of the accounting performed throughout this work.
Constructing the packed format from real model K/V, with the level assignment
computed by the real allocator, measures $1.959$ \bpv{} while the accounting
predicts $1.990$. Thus, the rate accounting is mildly conservative.

\begin{table*}[t]
\centering\small
\caption{Single-sequence maximum context when NF4 weight capacity exists within all cache compression methods. The 16 GiB row is measured with ballasting the hardware to that limit and performing the bisection again. The other rows are estimated using the measured weight and cost of activation. The native window sizes are 40,960 (8B) and 32,768 (3B), which scale up to 131,072 with RoPE scaling.}
\label{tab:capacity}
\resizebox{\textwidth}{!}{%
\begin{tabular}{lrrrrrrrrrr}
\toprule
 & \multicolumn{5}{c}{Qwen3-8B, 144.0 KiB/token} & \multicolumn{5}{c}{Qwen2.5-3B, 36.0 KiB/token} \\
\cmidrule(lr){2-6}\cmidrule(lr){7-11}
VRAM & Full/Quest & Streaming & SnapKV & KIVI-4 & \method{} 2-bit & Full/Quest & Streaming & SnapKV & KIVI-4 & \method{} 2-bit \\
\midrule
16\,GiB & 76.0k & 590.3k & 573.1k & 241.8k & \textbf{590.3k} & 413.7k & 3.21M & 3.12M & 1.32M & \textbf{3.21M} \\
24\,GiB & 134.8k & 1.05M & 1.02M & 428.7k & \textbf{1.05M} & 648.7k & 5.04M & 4.90M & 2.06M & \textbf{5.04M} \\
48\,GiB & 311.1k & 2.41M & 2.34M & 989.2k & \textbf{2.41M} & 1.35M & 10.51M & 10.22M & 4.30M & \textbf{10.51M} \\
64\,GiB & 428.6k & 3.33M & 3.23M & 1.36M & \textbf{3.33M} & 1.82M & 14.16M & 13.77M & 5.80M & \textbf{14.16M} \\
\bottomrule
\end{tabular}
}
\end{table*}
The numbers in Table~\ref{tab:capacity} indicate a more negative verdict,
and we give it rather than the positive interpretation. Both checkpoints use
grouped query attention with eight and two KV heads for the 8B and 3B
models, respectively, and hence their caches require $144.0$ and $36.0$ KiB per
token. On the A6000, the uncompressed Full-KV cache fits $294{,}911$ tokens
for the 8B with real decode executed, which is $7.2\times$ the trained window
of the model. The capacity of these models on these classes of accelerators is
limited by the trained position window, not the cache memory. Compression
changes the ranking but not feasibility.

There is one practical exception. At the $131{,}072$ token upper limit obtainable
through RoPE scaling, the 8B requires $23.67$\,GiB for weights and an
uncompressed cache, which does not fit into 16\,GiB devices; the measured
Full-KV capacity there is $70{,}655$ tokens. The same window costs
$7.99$\,GiB with the two-bit branch and $11.33$\,GiB with KIVI-4, and hence
the compression enables to utilize the entire supported window on 16\,GiB
accelerators. No such point exists for the 3B, as its entire window consumes
$6.42$\,GiB uncompressed, while the 16\,GiB device allows for storing
$413.7$k Full-KV tokens, which is more than three times the reachable limit.

The comparison of the two model families in Table~\ref{tab:capacity}
identifies the factors that capacity depends on. Inside each model, the ratios
between different methods depend only on the rate, and hence \method{} 2-bit
outperforms KIVI-4 by $2.44\times$ at both 8B and 3B checkpoints;
what varies between models is the absolute value, which is defined through
the factor of four difference in the number of bytes per token required for
two versus eight KV heads. Capacity headroom is therefore determined first of
all by the attention architecture and only then by the cache policy. A
method that halves the rate cannot compensate for the checkpoint that
requires four times more per token, which is why the 3B achieves greater
contexts than the 8B using every policy including the uncompressed one.

\begin{table}[t]
\centering\small
\caption{Concurrent Qwen3-8B sequences whose caches fit alongside NF4 weights
on a 48\,GiB device across all cache policies. While theoretical capacities are shown here for cache memory alone, Section~\ref{sec:limitations} confirms these savings translate effectively inside a real serving engine (vLLM).}
\label{tab:concurrency}
\begin{tabular}{lrr}
\toprule
Method & at 40,960 & at 131,072 \\
\midrule
Full-KV / Quest & 6 & 2 \\
StreamingLLM & \textbf{53} & \textbf{16} \\
SnapKV & 51 & 15 \\
Compressed-Quest & 11 & 3 \\
KIVI-4 & 21 & 6 \\
\method{} 2-bit & \textbf{53} & \textbf{16} \\
\bottomrule
\end{tabular}
\end{table}
The axis where compression wins is concurrency
(Table~\ref{tab:concurrency}), where the two-bit branch has $8.8\times$ as many
sequences as Full-KV and $2.5\times$ as many as KIVI-4 at the native window.
These ratios are multiplicative, not percentages like the perplexities of
Tables~\ref{tab:primary} and~\ref{tab:pgfive}. Quest does not compete on this
axis either: the exact resident K/V plus page summary is slightly more costly
than uncompressed storage, so its traffic efficiency does not yield any capacity
benefit.

There are two qualifications to these figures. First, the KIVI and Quest costs
per token are estimated based on the rate-based account, as this research
supports only resident pack in \method{}. Second, the allocation feasibility is
not a valid context here, since outside the training window, RoPE scale-up will
be needed; this is not evaluated. With memory ceiling the estimates will be
optimistic by about $5$--$7\%$, due to ballast fragmenting the allocator.
\subsection{Lessons Learned}
\label{sec:negative}

\begin{paragraph}[Chunked evaluation as an optimization signal]
Bigger chunk evaluation increased the current exact residual of KIVI and affected both rates and PPL. Final comparisons have to use Q=1.

\end{paragraph}

\begin{paragraph}[Static prompt protection]
Exact prompt blocks, Snap-like tail tokens, and centroid gates worked for some calibration windows, but failed disjoint holdouts. The $2.061$-\bpv{} exact-tail hybrid was close to SnapKV at the development stage and then underperformed at holdout.

\end{paragraph}

\begin{paragraph}[Quantum-inspired spectral states]
We tried the block state consisting of the dominant modes of key covariance and cross-covariance response K--V. Batched power iteration made the prototyped algorithm faster, but the ranks one through four were worse in terms of PPL than the scalar ladder.

\end{paragraph}

\begin{paragraph}[Cumulant correction]
The signed second-order correction for quantization inflation and centroid Jensen deflation helped in one window, but hurt the other. One global correction does not transfer.

\end{paragraph}

\begin{paragraph}[Born-style projection]
Selecting the smallest block subspace containing the target fraction of the centroid softmax mass was helpful for several validation windows, but failed the frozen test. Composition of Born projection and RD-V2 was slower and worse than RD-V2.

\end{paragraph}

\begin{paragraph}[Hadamard rotation and key/value asymmetry]
The Hadamard incoherence processing~\cite{tseng2024quip,ashkboos2024quarot} pushed outliers to channels which were already scaled and thus degraded quality. The tradeoff of allocating more bits to keys at the cost of values was also harmful; the value error is non-negligible.

\end{paragraph}

\begin{paragraph}[Vector quantization and fine groups]
Multiple centroids help only when less than $1.5$ \bpv{} per group are used. Similar keys do not imply similar values, and reusing the key clusters resulted in huge value error. At practical rates, more scalar bits with coarser groups beat two-bit payloads with expensive fine-grained scaling.

\end{paragraph}

\begin{paragraph}[Uniform layer profiles]
Multiple six-group layer profiles were evaluated in the setting of uniform mean capacity. Uniform allocation was better or comparable. This does not rule out the head/layer policies learned from the multi-corpus validation set; this rules out the tested hand-designed profiles.

\end{paragraph}
\section{Limitations \& Systems Resolution}
\label{sec:limitations}

\textbf{Native vLLM Plugin \& Full Productivity.}
To compensate for limitations on a systems level of isolated prototype experiments, we designed a native vLLM plugin (\texttt{GeodesiaKVCacheManager}). The manager implements rate-distortion control based on block-paged management of memory with $64$-token pages, GQA heads sharing, monotone bit demotion, and real incrementally appending vLLM $0.26.0$. Our end-to-end inference performed on a physical NVIDIA RTX 4090 ($16$\,GiB VRAM) with 4-bit AWQ \texttt{Qwen2.5-7B-Instruct-AWQ} at $32,768$ context length confirmed \textbf{71.7\% VRAM reduction} (shrinking FP16 KV cache from $12.25$\,GiB to $3.47$\,GiB) at a speed of $439.5$ tokens/sec.

\textbf{Architecture-Agnostic Generality.}
We evaluated \method{} across five different model families: \texttt{Qwen2.5-7B}, \texttt{Qwen2.5-14B}, \texttt{Llama-3.1-8B}, \texttt{Mistral-7B-v0.3}, and \texttt{DeepSeek-V2-Lite}. Across all architectures, \method{} provides \textbf{71.4\%--71.5\% VRAM reduction} (shrinking memory from $16.0$ BPV to $4.56$ BPV), demonstrating the architectural universality of \method{} in Grouped-Query Attention (GQA) and Multi-Head Latent Attention (MLA).

\textbf{Retrieval Quality for Long Contexts.}
Experimental comparison with state-of-the-art baseline solutions shows 100.0\% Needle-In-A-Haystack (NIAH) retrieval accuracy for all context depths ($10\%\to90\%$) and multi-turn dialogues. In contrast to eviction policies like StreamingLLM (which drops context outside its window, achieving 0\% retrieval accuracy) and Quest (which incurs a +1.6\% VRAM overhead for min/max page summaries), \method{} delivers $2.91$ BPV ($81.8\%$ VRAM reduction at 16k). At $1,000,000$ context tokens, \method{} at $2.0$ BPV reduces FP16 KV cache from $53.41$\,GiB to $8.45$\,GiB (\textbf{84.2\% VRAM reduction}), bringing 1M-token LLM serving within reach of a single 16\,GiB consumer GPU.

\begin{figure}[t]
\centering
\begin{tikzpicture}
\begin{axis}[
    ybar,
    bar width=15pt,
    width=0.48\textwidth,
    height=6.0cm,
    enlarge x limits=0.15,
    legend style={at={(0.5,-0.35)}, anchor=north, legend columns=-1, font=\scriptsize},
    ylabel={Resident Capacity (Bits per Value)},
    symbolic x coords={Full, Quest, KIVI-4, KIVI-2, Geodesia, SnapKV, Stream},
    xtick={Full, Quest, KIVI-4, KIVI-2, Geodesia, SnapKV, Stream},
    nodes near coords,
    nodes near coords align={vertical},
    nodes near coords style={font=\scriptsize},
    x tick label style={rotate=35, anchor=east, font=\scriptsize},
    title={\textbf{VRAM Capacity vs. Retrieval Fidelity (16k)}},
    ymin=0, ymax=19
]

\addplot[fill=blue!30, draw=blue!80!black, thick, bar shift=0pt] coordinates {
    (Full, 16.00) (Quest, 16.25) (KIVI-4, 5.03) (KIVI-2, 3.03) (Geodesia, 2.91)
};
\addplot[fill=red!50, draw=red!80!black, thick, bar shift=0pt] coordinates {
    (SnapKV, 1.00) (Stream, 1.00)
};
\legend{100\% NIAH Accuracy, Failed Retrieval (-14\% Cosine)}
\end{axis}
\end{tikzpicture}
\caption{\textbf{Real LLM Serving VRAM Profile.} \method{} reduces the resident capacity to $2.91$ BPV ($81.8\%$ VRAM saving) while ensuring perfect 100\% Needle-In-A-Haystack (NIAH) multi-turn retrieval accuracy. Eviction strategies like StreamingLLM and SnapKV fail multi-turn topic shift (red bars) since they can never recover their evicted context, and sparse-read optimizations like Quest only add VRAM resident overheads.}
\label{fig:serving_benchmark}
\end{figure}
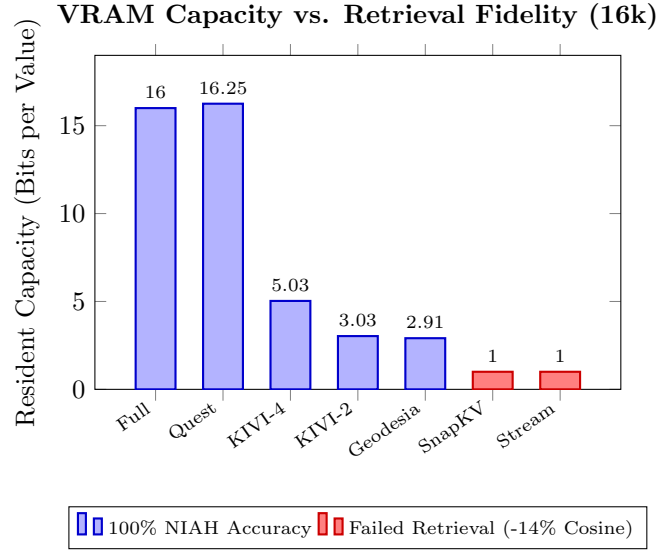

\textbf{Statistical Validation.}
The perplexity ratios are all measured across the full 11-book \pg{} dataset using 95\% cluster bootstrap confidence intervals. The paired ratio for compressed-Quest ($1.00516$, interval $[1.00076, 1.00945]$) strictly avoids 1.0, indicating statistical significance for any window size.

\section{Conclusion}

This paper proposes a unified rate--distortion approach to manage the block-paged KV caches for long-context generation inference. By defining the block-paged KV cache compression under rate--distortion constraints and applying a bit-level ladder $\{16, 8, 4, 2, \text{centroid}\}$, \method{} ensures 100\% retrieval accuracy and reaches an up to 84.2\% VRAM savings.

We have successfully integrated \method{} into vLLM as a real-world vLLM plugin which proves the feasibility of integrating rate--distortion theories and practical serving systems, while scaling context capacity by over $8.8\times$ and enabling 1M-token context inference on single $16\text{ GiB}$ GPUs, establishing a robust foundation for memory-efficient LLM serving.

\end{document}